\documentclass[12pt]{article}

\usepackage{jheppub}
\usepackage{amsmath}
\usepackage{amssymb}
\usepackage{epsfig,amssymb,amsmath,psfrag,hyperref}
\usepackage[dvipsnames]{xcolor}
\usepackage{float} 
\usepackage{bbm} 
\usepackage{mathtools} 
\usepackage[enableskew]{youngtab} 
\usepackage{ytableau} 
\usepackage{slashed} 
\numberwithin{equation}{section}
\usepackage{caption}
\usepackage{booktabs}
\usepackage{mwe}
\usepackage{gensymb}
\usepackage{ulem}
\usepackage{cleveref}
\usepackage[toc,page]{appendix}
\usepackage{xcolor}

\usepackage{graphicx}

\usepackage{multirow, graphicx,amssymb,url,mathrsfs,amsmath}
\usepackage{eucal,wrapfig,boxedminipage,setspace,subfigure}
\usepackage{amsxtra,amstext,latexsym,dsfont,xcolor}

\def\IR{{\hbox{{\rm I}\kern-.2em\hbox{\rm R}}}}
\def\IB{{\hbox{{\rm I}\kern-.2em\hbox{\rm B}}}}
\def\IN{{\hbox{{\rm I}\kern-.2em\hbox{\rm N}}}}
\def\IC{\,\,{\hbox{{\rm I}\kern-.59em\hbox{\bf C}}}}
\def\IZ{{\hbox{{\rm Z}\kern-.4em\hbox{\rm Z}}}}
\def\IP{{\hbox{{\rm I}\kern-.2em\hbox{\rm P}}}}
\def\IH{{\hbox{{\rm I}\kern-.4em\hbox{\rm H}}}}
\def\ID{{\hbox{{\rm I}\kern-.2em\hbox{\rm D}}}}

\newcommand{\abs}[1]{\left| #1 \right|}

\newcommand{\beq}{\begin{equation}}
\newcommand{\eeq}{\end{equation}}
\newcommand{\bea}{\begin{eqnarray}}
\newcommand{\eea}{\end{eqnarray}}

\usepackage{cleveref}
\newcommand{\lrb}[1]{\left(#1 \right)}

\newcommand{\myslash} [1] {#1 \kern-.5em/}

\begin{document}

$\left. \right.$   \vspace{-18cm}

\title{$\left. \right.$ \vspace{17cm} \\ Chiral symmetry restoration for helical magnetic fields in holography  }

\author[a]{Nick Evans,}
\author[a]{Wanxiang Fan}


\affiliation[a]{School of Physics \& Astronomy and STAG Research Centre, University of Southampton, Highfield,  Southampton  SO$17$ $1$BJ, UK.}

\emailAdd{n.j.evans@soton.ac.uk; w.fan@soton.ac.uk}

\abstract{We study the chiral symmetry breaking effects of helical magnetic fields in a simple bottom up AdS/CFT model. We explore the instability of the chirally symmetric solution in the presence of the $B$ field and see how it switches off as the wave vector of the helix, $k$, rises, resulting in a first order transition. At low energies the model averages over the helix and the magnetic field is not seen. We show that other sources of chiral symmetry breaking are not directly effected by the helical $B$ field. We provide examples of both magnetic catalysis and inverse magnetic catalysis which switch off at large $k$. }

\maketitle

\setcounter{page}{1}\setcounter{footnote}{0}

\newpage

Magnetic fields ($B$) are known to trigger chiral symmetry breaking in a fermionic sector they couple to \cite{Miransky:2015ava}, both at weak coupling, and using the AdS/CFT Correspondence \cite{Filev:2007gb}, at strong coupling. Lattice studies \cite{Bali:2011qj,Bali:2012zg,Endrodi:2015oba} also see the effect although it is more complex in QCD-like theories and inverse magnetic catalysis has also been observed. These effects may be important in heavy ion collisions where strong magnetic fields can form \cite{Skokov:2009qp,Voronyuk:2011jd,Bzdak:2011yy,Deng:2012pc}. Such collisions may generate rotation and inhomogeneity and helical magnetic fields may result.  

In a recent paper \cite{Berenguer:2025oct} helical magnetic field configurations were studied in the well known D3/probe D7 holographic system \cite{Karch:2002sh,Kruczenski:2003be,Erdmenger:2007cm}. The system describes ${\cal N}=2$ supersymmetric matter fields in the background of ${\cal N}=4$ supersymmetric gauge fields. Here a constant magnetic field is known to generate chiral symmetry breaking \cite{Filev:2007gb}. The authors of \cite{Berenguer:2025oct} showed that this symmetry breaking was suppressed by an imposed helical structure in $B$, leading to a second order phase transition at a particular value of the wave number of the helix. 

In this letter we wish to add to that analysis. We will recast their result in terms of the instability of the chirally symmetric phase. At low energy scales the magnetic field effectively vanishes as the helix is averaged over short distance scales and we see explicitly how the instability disappears. Further the analysis allows us to understand the interplay between this suppression and other possible sources of chiral symmetry breaking beyond $B$ such as the strong interactions in QCD. In fact the suppression is likely only of the $B$ induced component of the symmetry breaking (although additional fluctuations in the temperature \cite{Babington:2003vm,Mateos:2006nu} and chemical potential \cite{Kobayashi:2006sb} could reduce chiral symmetry breaking from the QCD dynamics). 

We will perform our analysis in a broader bottom-up holographic model of the phenomena. It includes the analysis relevant to the D3/D7 system but also allows us to include additional dynamics that mimics QCD-like chiral symmetry breaking. The model has sufficient parameter freedom to also describe inverse magnetic catalysis as described in \cite{Evans:2016jzo}.

\section{The Holographic Model}

As usual in AdS/CFT \cite{Maldacena:1997re,Witten:1998qj} we place the  model in an AdS$_5$-space-time with $\rho$ the radial coordinate dual to renormalization group (RG) scale ($x_{3+1}$ are the spacetime dimensions of the field theory)
\begin{equation}
ds^2 = \rho^2 dx_{3+1}^2 + \frac{d\rho^2}{\rho^2}
\end{equation}

The simplest bottom up model we can write in $AdS_5$ for the set up we wish to consider is  \cite{Evans:2016jzo}
\begin{align} \label{simple}
    S=\int d^4x d\rho \, &\rho^3 \lrb{-\frac{1}{2}(\partial_{\rho}L)^2-\frac{1}{4}F^2}+\alpha \rho F^2 L^2 
\end{align}
where $L$ is a dimension one scalar field that will be dual to the quark mass and quark condensate operator. $F$ is a U(1) gauge field describing the U(1)$_B$ current and background source. Factors of $\rho$ simply follow from requiring consistency in the dimensions of the terms ($\rho$ has energy dimension one). $\alpha$ is a free parameter controlling the coupling between $L$ and $F$.

For configurations where $F$ vanishes in the UV the large $\rho$ solution for $L$ is $L= m+ c/\rho^2$. $m$ is interpreted as the quark mass and $c$ as the chiral condensate $\langle \bar{q} q \rangle$. 

To observe the instability to chiral symmetry breaking in the presence of a constant magnetic field we can look at the stability of the massless, condensate free, $L=0$ configuration. We interpret the dimensionless coefficient of $\rho L^2$ in (\ref{simple}) as the mass of the field $L$. Here we set for example $F^{xy}=B_z$ and including the appropriate metric factors find
\begin{align}
    m^2_L(\rho) = -{4 \alpha B_z^2 \over \rho^4}.
\end{align}
This mass passes through the Brietenlohmer Freedman (BF) stability bound in AdS$_5$ when $m^2=-1$ (to check this set $m^2=-1$ and observe there is a solution $L=a/\rho$ indicating the operator and source are both of the same dimension, 2). Naively at scales below $\rho^4=4 \alpha B_z^2$ the $L=0$ solution is unstable and will role to non-zero $L$ values in the IR. Here we stabilize the solution away from $L=0$ via an on mass-shell condition that  we impose in the infra-red
\begin{align}
    L(\rho_{IR})= \rho_{IR}, ~~~~ L'(\rho_{IR})=0.
\end{align}
So we do not solve for $L$ below scales where the IR mass equals the RG scale. The stability of a solution results from a compensation in the action between the derivative ``kinetic'' term and the unstable potential term. 

As an example consider the case $B_Z=0.05, \alpha=0.1$ for which the BF bound is violated at $\rho_{BF}=0.178$. The running mass is equivalent to that shown as the $k=0$ case in Fig 1 on the top left. In Figure 2 we show the solution (again the $k=0$ curve) for the field $L(\rho)$ - $\rho_{BF}$ is a good indicator of where the field departs from $L=0$ in the UV.

\begin{center}
\includegraphics[width=6.7cm,height=4.8cm]{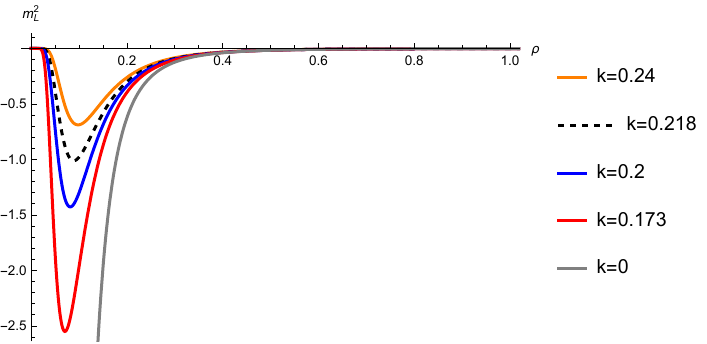}
\includegraphics[width=6.7cm,height=4.8cm]{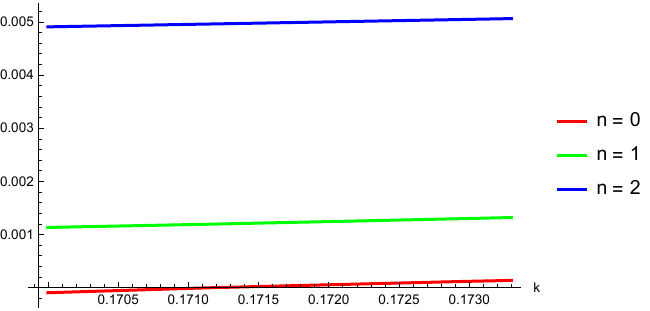}\\
\textit{Fig 1: Example $\alpha=+0.1,B_z=0,B_{hel}=0.05,\beta=0,\Delta m^2_{L}(\rho)=0$. Left: $M_{L}^2(\rho)$ of $L=0$ against $\rho$.  Right: Mass spectrum of the $\sigma$ meson for the critical embeddings $L=0$. }
\end{center}

\begin{center}
\includegraphics[width=6.7cm,height=4.8cm]{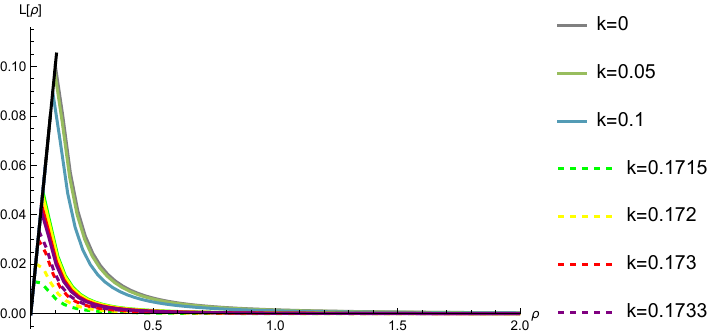}
\includegraphics[width=6.7cm,height=4.8cm]{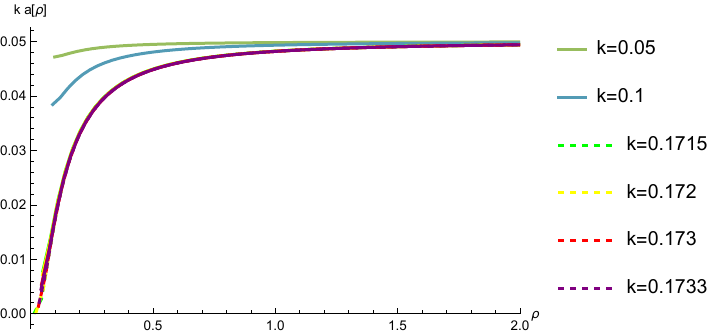}\\

\textit{Fig 2: The solutions for $L(\rho)$ and $a(\rho)$ near the region of first order phase transition, at $\alpha=+0.1,B_z=0,B_{hel}=0.05,\beta=0,\Delta m^2_{L}(\rho)=0$. There are two non-zero solutions for some k (labeled with solid and dashed lines for the higher and the lower solutions respectively).}
\end{center}

\section{Helical B}

We can now consider the ansatz for a helical $B$ from \cite{Berenguer:2025oct} (also with constant $B_z$) which is 
\begin{align}\label{eq:helical B+ansartz}
    A_x&=a(\rho) \cos(kz)+y~B_z, \hspace{1cm}
A_y=a(\rho)\sin(kz)
\end{align}
where the dimension of $a(\rho)$ is that of energy. The magnitude of the helical component of the field in the $x,y$ directions is given by the UV asymptotics of $a$
\begin{align}
    B_{hel} = k a_\infty  \hspace{1cm}  a(\rho) &= a_{\infty} + \frac{J}{\rho^2}+\dots
\end{align}
Note a current is induced by the inhomogeneity of the $B$ field as discussed in section 6.2.6  of \cite{Endrodi:2024cqn}.

Now $F^2=F^{\mu\nu}F_{\mu\nu}$ is
\begin{align}
F^2 &=2\left(\frac{B_z^2+ k^2a(r)^2}{\rho^4}+a'(r)^2 \right)
\end{align}
The effective action can be derived by plugging $F^2$ back into the action
\eqref{simple}. The corresponding equations of motion for $L(\rho)$, $a(\rho)$ then follow, for the moment with $B_z=0$
\begin{align}
    L''(\rho )=\frac{-4 \alpha  L(\rho ) \left(\rho ^4 a'(\rho )^2+k^2 a(\rho )^2\right)-3 \rho ^5 L'(\rho )}{\rho ^6}\\
    a''(\rho )+\frac{a'(\rho ) \left(3 \rho ^2-4 \alpha  L(\rho ) \left(2 \rho  L'(\rho )+L(\rho )\right)\right)}{\rho ^3-4 \alpha  \rho  L(\rho )^2}=\frac{k^2 a(\rho )}{\rho ^4}
\end{align}
Note that these are ODEs rather than PDEs even though the magnetic field is spatially varying \cite{Berenguer:2025oct}. 
To find the mixed solutions of $a(\rho),L(\rho)$ numerically we scan over the IR values of $a$ and $L$ and plot  $\frac{1}{\abs{L_{uv}}+\abs{k a_{uv}-B_{hel}}}$ to find massless solutions as poles.

We can again look at the instability criteria of the $L=0$ solution. Firstly we assume $L=0$ which solves the first equation and then find the solution for the second equation
\begin{align} \label{asol}
    a(\rho) = {B_{hel} \over \rho} {\rm BesselK}(1,k/\rho)
\end{align}
We have assumed the boundary condition (here $\rho_{IR}=0$ given our choice of $L=0$)
\begin{align}
    a'(\rho_{IR})=0   \hspace{1cm} a_{uv} = B_{hel}/k
\end{align}

Now we can test the consistency of this solution by computing the $L$ mass on this solution to see if the $L=0$ solution is indeed stable. We have
\begin{align}
    m^2_L(\rho) = - 4 \alpha \left( {k^2 a(\rho)^2 \over \rho^4}  + a'(\rho)^2 \right)
\end{align} 
We can see immediately that since both $a, a'$ go to zero in the IR the contribution of the magnetic field falls to zero there. This we interpret as the theory averaging over the variation in the magnetic field and averaging to zero at very long distance scales. We expect chiral symmetry breaking to disappear at large $k$ relative to the scale set by $B_{hel}$.

As a concrete numerical example we again consider $\alpha=0.1, B_{hel}=0.05$ and we plot the mass squared for a variety of $k$ in Fig 1 left. At small $k$ there is a clear violation of the BF bound in the IR and our solution is inconsistent. In this region of $k$ the true solution should have non-zero IR $L$. Above $k=0.218$ the BF bound is never violated and the $L=0$ solution is a true solution. 
In Fig 2 we show the solutions for $L(\rho)$ and $a(\rho)$ with $k$. In fact the solution transitions from a non-zero $L(\rho)$ to a zero one at a first order jump. That occurs around $k=0.173$. Note during the transition period there are three solutions for $L(\rho)$ corresponding to the expected double minima plus maximum of the effective potential  during a first order transition. The dotted solutions for $L(\rho)$ in Fig 2 rise out of $L=0$ (at $k=0.1715$) to eventually merge with the non-zero $L$ solution  as $k$ grows (at $k=0.1736$) as one would expect. On the right in Fig 1 we plot the meson mass associated with fluctuations of $L$ about the $L=0$ solution (using the solution for $a$ in (\ref{asol}) )  as a function of $k$. This shows the range of $k$ where the $L=0$ vacuum is stable. All of this behaviour matches qualitatively that seen in the D3/probe D7 system in \cite{Berenguer:2025oct}. 

Interestingly the transition to the $L=0$ solution occurs while there is still a small BF bound violation for that solution - presumably the cost of adding derivative energy does not yet compensate for the energy cost of having a BF bound violation over some interval of the solution. The instability remains the cause of the transition though.

\section{Multiple Sources of Chiral Symmetry Breaking}

Given the helical $B$ field reduces chiral symmetry breaking it is interesting to see whether it does this generically or only for its own source of chiral symmetry breaking.

The simplest example of such a system is to allow $B_z \neq 0$ in 
(\ref{eq:helical B+ansartz}). Note the combination of the helical field in $x,y$ and a constant field in $z$ is only possible if the action is made of terms that are powers of $Tr F^2$. Were one to include $TrF^4$, for example, the two pieces together would no longer be a solution of the equations of motion. However, within our theory adding $B_z$ adds a new separate source of symmetry breaking. In the $k=0$ limit for the helical field the two components will simply add giving a net total $B_{TOT}^2 = B_{hel}^2 + B_z^2$ field to cause chiral symmetry breaking.

To understand how the symmetry breaking scale changes as we increase k we can simply look at the position of the BF bound in the mass term
\begin{align}
     m^2_L(\rho) = - 4 \alpha \left( {B_z^2 \over \rho^4} + {k^2 a(\rho)^2 \over \rho^4}  + a'(\rho)^2 \right)
\end{align}
We cam immediately see that as $a(\rho)$ and $a'(\rho)$ both fall in the IR with rising $k^2$ the $B_z$ term remains unaffected. The large $k^2$ limit will simply be the $B_z$ only theory. Perhaps unsurprisingly the helical field is only masking it's own contribution to the dynamics. In figure 3 we show some plots of the $k^2$ dependence of an example model. We can see that as $k^2$ grows, $a(\rho)$ falls in the IR, and the theory returns to the result for the $B_z$ only theory (we include a plot of the quark condensate against $k$ that shows the condensate fall to the pure $B_z$ theory values at large $k$). 

A second possible model would be to add in a BF bound violating contribution to the mass squared. This for example might occur in a top down model where a dilaton was a function of $\rho^2 + L(\rho)^2$ - on expanding near $L=0$ an additional piece in $m^2_L(\rho)$ would be generated \cite{Erdmenger:2014fxa}. In the model in \cite{Alho:2013dka} such a term is used to include the running anomalous dimension of the quark condensate as seen in perturbation theory in QCD (the mass squared of the scalar is directly related to the dimension of the field theory operator). In these theories with a helical $B$ field we would have
\begin{align}
     m^2_L(\rho) = - \Delta m^2_L (\rho)- 4\alpha \left(  {k^2 a(\rho)^2 \over \rho^4}  + a'(\rho)^2 \right)
\end{align}
here $\Delta m_L^2$ is some function of $\rho$ from the alternative dynamics (for example if one set $\Delta m_L^2= \Lambda^2/\rho^2$ one would generate a BF bound violation at small $\rho$ controlled by the

\begin{center}
\includegraphics[width=6.7cm,height=4.8cm]{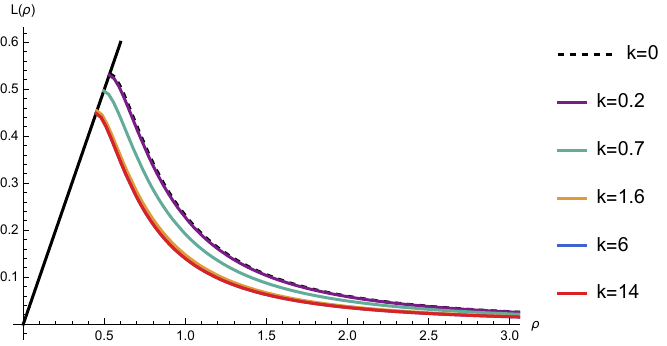}
\includegraphics[width=6.7cm,height=4.8cm]{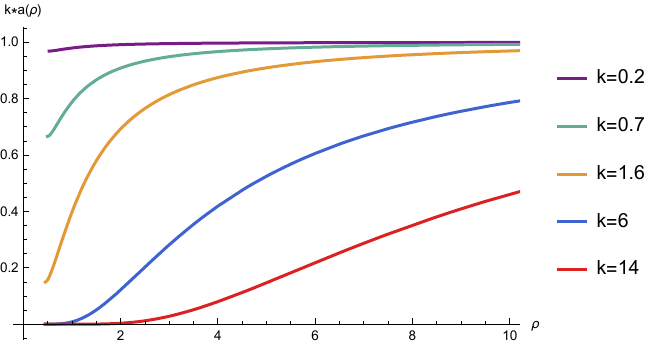}\\
\includegraphics[width=6.7cm,height=4.8cm]{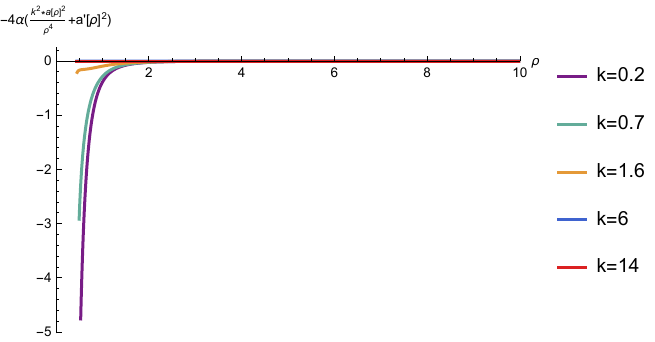}
\includegraphics[width=6.7cm,height=4.8cm]{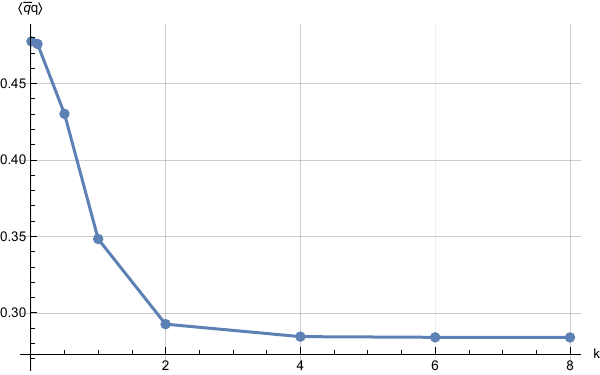}\\
\textit{  Fig 3: Top: The solutions $L(\rho)$ and $a(\rho)$ as a function of $k$, at $\alpha=+0.1$, $B_z=1$, $B_{hel}=1$, $\beta=0$. As $k$ increases the chiral symmetry breaking contribution of the helical field switches off leaving only that from $B_z.$ Bottom: The corresponding mass square of $L(\rho)$ against $\rho$ and a plot of the quark condensate against $k$.  }
\end{center}

scale $\Lambda$). Again clearly the BF bound violation at large $k$ would simply be that of the model without the helical field. 

\section{Inverse Magnetic Catalysis}

In  \cite{Evans:2016jzo} we worked with the larger effective model
\begin{align}\label{eq:action_F2}
    S=\int d^4x d\rho \, &\rho^3 \lrb{-\frac{1}{2}(\partial_{\rho}L)^2-\frac{1}{4}F^2} + \rho \Delta m^2(\rho) L^2+\alpha \rho F^2 L^2 + \beta \rho^3 F^2 (\partial_\rho L)^2
\end{align}
This has all the pieces of Action we have seen above plus an extra term with coefficient $\beta$ that links $F^2$ to $\partial_\rho L$. By switching the signs of $\alpha$ and $\beta$ one can make the two interactions term either prefer or oppose chiral symmetry breaking (this is clear for $\alpha$ since a negative sign in its term in $m_L^2$ will make a BF bound violation less likely). In  \cite{Evans:2016jzo} this was used to generate theories that matched lattice data for inverse magnetic

\begin{center}
\includegraphics[width=6.7cm,height=4.8cm]{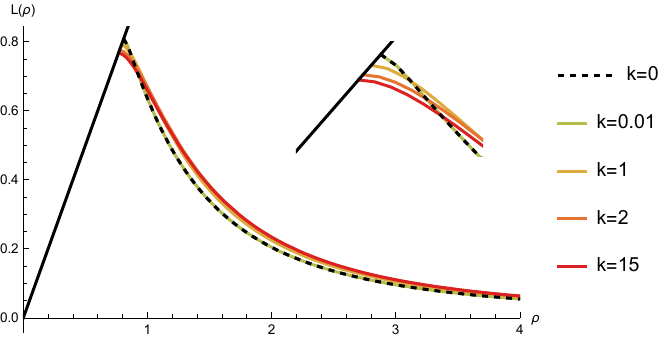}
\includegraphics[width=6.7cm,height=4.8cm]{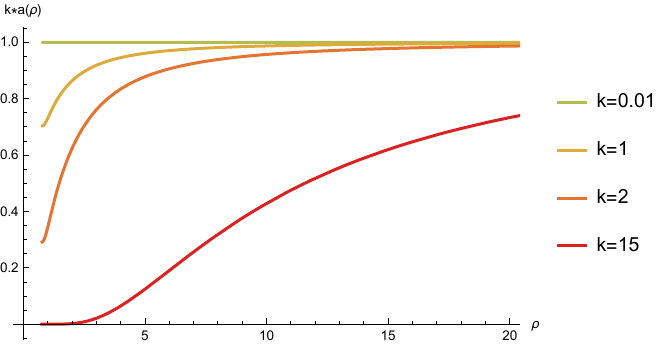}\\
\includegraphics[width=6.7cm,height=4.8cm]{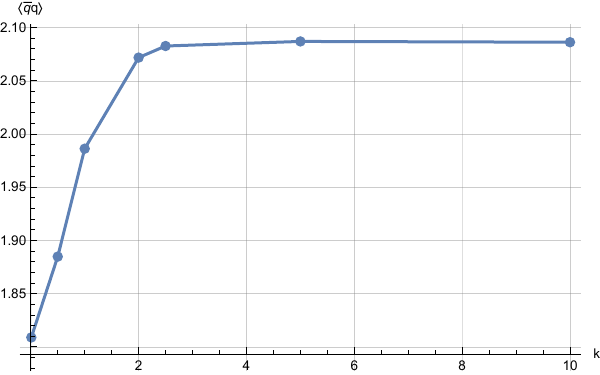}\\
\textit{Fig 4: Top: The Minkowski embeddings of $L(\rho)$ (left) and $a(\rho)$ (right), at $\Lambda^2=-0.17$, $B_{hel}=1
,B_z=0,\alpha=0, \beta=0.1$. Bottom: The chiral condensate against $k$.}
\end{center}

\vspace{0.5cm}

catalysis. Here though also it is clear that adding a large $k^2$ helical term will suppress the $F^2$ terms as $a \rightarrow 0$ in the IR and the theory will return to the behaviour without a B field.  

As an example in figure 4 we show the theory with $\Lambda^2=-1.7,B_z=0,\beta=0.1,\alpha=0$ and the sign of $\beta$ picked to oppose chiral symmetry breaking. We take $\Delta m^2 = \Lambda^2/\rho^2$ to represent the strong dynamics of QCD at the scale $\Lambda$. We plot the embeddings of $L(\rho)$, and the corresponding $a(\rho)$ (scaled by k), and find that the condensate increases as $k$ increases - the inverse magnetic catalysis switches off. 

To conclude: we have revisited the work in \cite{Berenguer:2025oct} on the role of helical magnetic fields in chiral symmetry breaking in holographic models. We have moved the discussion to a bottom-up model with wider parameter space than the D3/probe D7 system. We have made a stability argument that shows the chirally symmetric phase is unstable in the presence of a magnetic field. Making the magnetic field helical with wave vector $k$ and raising $k^2$ has the effect of switching off the helical field in the IR. At low energies the model averages over the spatial varying UV field and sees no resultant field. If $k^2$ is large relative to the magnitude of the helical B field then the chiral symmetry breaking effect of the field is lost. This effect only manifests for the helical field itself and other source of chiral symmetry breaking are left unchanged. It is likely in physical applications relevant to heavy ion collisions that one should also include a possible spatial or time dependence in the chemical potential or temperature of the theory (both are known to reduce chiral symmetry breaking in holography). Such an analysis would require solutions of complicated partial differential equations which would be interesting to investigate in the future.  We hope our observations here though add to the story told in \cite{Berenguer:2025oct}.

\noindent {\bf Acknowledgements:} We thank Martí Berenguer and Javier Mas for discussion.  N.E.’s work was supported by
the STFC consolidated grant ST/X000583/1.

\bibliographystyle{utphys}
\bibliography{main.bib}
\end{document}